\documentclass[apj]{emulateapj}

\usepackage{graphicx,natbib,color}

\shorttitle{The sub-arcsecond hard X-ray structure of loop footpoints in a solar flare}
\shortauthors{Kontar {\it et al.}}

\begin{document}


\title{The sub-arcsecond hard X-ray structure of loop footpoints in a solar flare}


\author{E. P. Kontar, I. G. Hannah, N. L. S. Jeffrey, and M. Battaglia}
\affil{Department of Physics and Astronomy,
University of Glasgow, G12 8QQ, United Kingdom}




\begin{abstract}
The newly developed X-ray visibility forward fitting technique
is applied to Reuven Ramaty High Energy Solar
Spectroscopic Imager (RHESSI) data of a limb flare to investigate the energy
and height dependence on sizes, shapes, and position of hard X-ray chromospheric
footpoint sources. This provides information about the electron transport
and chromospheric density structure. The spatial distribution of
two footpoint X-ray sources is analyzed using PIXON, Maximum Entropy Method, CLEAN
and visibility forward fit algorithms at nonthermal energies
from $\sim 20$ to $\sim 200$ keV. We report, for the first time, the vertical extents
and widths of hard X-ray chromospheric sources measured as a function
of energy for a limb event. Our observations suggest that both the vertical
and horizontal sizes of footpoints are decreasing with energy.
Higher energy emission originates progressively deeper in the chromosphere
consistent with downward flare accelerated streaming electrons. The ellipticity
of the footpoints grows with energy from $\sim 0.5$ at $ \sim 20$ keV to $\sim 0.9$ at
$\sim 150$ keV. The positions of X-ray emission are in agreement with an exponential
density profile of scale height $\sim 150$~km. The characteristic size
of the hard X-ray footpoint source along the limb is decreasing with energy suggesting a
converging magnetic field in the footpoint. The vertical sizes
of X-ray sources are inconsistent with simple collisional
transport in a single density scale height but can be explained using
a multi-threaded density structure in the chromosphere.

\end{abstract}



\section{Introduction}

In the standard flare scenario, electrons that were accelerated
in the corona stream downwards toward the dense layers of the solar atmosphere,
where they are stopped via collisions producing
intense hard X-ray (HXR) emission in the chromosphere. Higher energy electrons
penetrate deeper into the chromosphere. Therefore, measurements of the spatial
structure of the HXR emission as a function of energy provide information about the
chromospheric density. Being optically thin, X-rays give the most direct information
about the spatial and energy distribution of energetic electrons in the solar
atmosphere. Prior to the launch of the Ramaty High Energy Solar Spectroscopic Imager
(RHESSI) \citep{lin2002}, HXR instruments typically had limited imaging-spectroscopy
capabilities \citep{Kosugi1992yohkoh}. RHESSI's aptitude to image in various energy
ranges opens new horizons for studying the detailed structure of HXR emitting
sources in the chromosphere.

RHESSI does not directly image the Sun but uses 9 pairs of Rotating Modulation
Collimators (RMCs) to time-modulate spatial information in the signal obtained in
its germanium detectors \citep{hurford2002}. Each RMC has a different thickness of its slits and
slats making it sensitive to different spatial scales, providing modulation at nine
spatial frequencies. The reconstruction of an image from these time modulated lightcurves,
can be accomplished by various imaging algorithms
\cite[][]{emslie2003,BattagliaBenz2007,KruckerLin2008,Saint-Hilaire_etal2008,dennis2009}.
The new visibility based approach to RHESSI imaging starts by summing
(stacking) the lightcurves per roll bins over a few spin periods
of the spacecraft \citep{schmahl2007}. The fitted amplitudes and the phases
in the individual roll bins are X-ray visibilities.
This effectively provides two dimensional spatial Fourier
components (X-ray visibilities) over a wide range of energies
\citep{hurford2002,schmahl2007}. To convert the time-modulated signal
or X-ray visibilities to an image is an inverse problem \citep[e.g.][]{Piana_etal2007,Prato_etal2009}.
The reconstructed images face unavoidable difficulties due to measurement errors,
finite coverage in Fourier space, and ill-posedness of the reconstruction problem.
The resulting reconstruction errors and small dynamic range makes it difficult
to accurately measure source sizes from reconstructed images. At best the imaging
resolution is down to 2 arcseconds but in practice is around 7 arcseconds
for the typical flare nonthermal energy range.

However, the moments of X-ray source distribution (source position, source size, etc)
can be inferred with higher precision either from the time-modulated signal
\cite[][]{aschwanden2002,KruckerLin2008,Saint-Hilaire_etal2008,Fivian_etal2009}
or visibilities \citep{Xu_etal2008,Kontar_etal08,dennis2009,Prato_etal2009}.
Thus, RHESSI measurements of X-ray source positions  can recover sub-arcsecond
information using RHESSI modulated lightcurves \citep{aschwanden2002,Liu_etal2006,Mrozek06}
or visibilities \citep{Kontar_etal08,dennis2009,Prato_etal2009}. This has allowed clear demonstration
of the height-energy dependence of HXR sources: higher energy sources originate at lower heights
\citep{aschwanden2002,brown2002,Liu_etal2006}. This has substantially improved
upon previous results with Yohkoh/HXT \citep{1992PASJ...44L..89M}. The recently
developed visibility-based technique \citep{schmahl2007}, allowed
\citet{Kontar_etal08} to improve previous measurements and infer characteristic sizes
(FWHM) of HXR footpoints and not just the centroid height. From this the convergence
of the magnetic flux and neutral hydrogen density distribution in the chromosphere
was inferred.

In this paper we study the structure of solar flare hard X-ray sources using various
imaging algorithms: PIXON, MEM, CLEAN and visibility forward fit.
Using RHESSI X-ray visibilities we find the characteristic
shapes and positions for different energy ranges.
We show that the technique of forward fitting X-ray visibilities allows us to determine
not only the FWHM of the sources but vertical and horizontal sizes of the sources,
which is required for examining the density structures of the chromosphere.
Theoretical relationships were compared with observations to find the density
structure of the chromosphere. The vertical size of the X-ray sources
is found to be larger than the ones predicted by a hydrostatic
atmosphere in thick-target scenario.
However, assuming that the electrons are propagating along
several narrow threads with different density profiles can explain
the measured vertical sizes of the sources.

\section{X-ray visibilities and characteristic sizes}

The spatial information about an X-ray source measured by RHESSI
for a given energy range and time interval
can be presented \citep{hurford2002,schmahl2007}
as two dimensional Fourier components or X-ray visibilities
\begin{equation}
V (u,v; \epsilon ) = \int _x \int _y I(x,y; \epsilon)e^{ 2\pi i (xu+yv)}dxdy
\label{eq:vis}
\end{equation}
where $I(x,y; \epsilon)$ is the observed image
at photon energy $\epsilon$. Then, reconstructed X-ray image
$I(x,y; \epsilon)$ is the inverse Fourier
transformation of measured X-ray visibilities $V (u,v; \epsilon )$.
Each of the nine RHESSI Rotating Modulating Collimators
(RMC) measures $V(u,v; \epsilon)$ at a fixed spatial frequency
(or a circle in the $(u,v)$ plane) corresponding to its angular
resolution and with a position
angle along the circles, which varies continuously
as the spacecraft rotates. Nine detector grids
with angular resolutions growing with detector number
are logarithmically spaced in the $(u,v)$ plane.
Since the measured visibilities sparsely populate the $(u,v)$
plane and have statistical uncertainties, the
direct inverse Fourier transform is impractical
\citep{hurford2002,schmahl2007,Massone_etal2009}
and alternative methods should be used.

Assuming a characteristic shape of X-ray source,
one can find the position and characteristic sizes directly
by fitting a 2D Fourier image of the model to the RHESSI visibilities.
Here, we assume that the sources can be presented as elliptical
Gaussian sources
\begin{equation}
I(x,y; \epsilon)=\frac{I_0(\epsilon)}{2\pi \sigma_x\sigma_y}\exp\left(-\frac{(x-x_0(\epsilon))^2}{2\sigma_x^2}-\frac{(y-y_0(\epsilon))^2}{2\sigma_y^2}\right),
\label{eq:Imodel}
\end{equation}
where $2\sqrt{2\ln2}\sigma_x$ and $2\sqrt{2\ln2}\sigma_y$ are FWHMs
of an elliptical Gaussian source in $x$ and $y$ direction respectively,
$x_0(\epsilon)$, $y_0(\epsilon)$ is the position
of the source, and $I_0$ is the total photon flux of the source.
One major advantage of the visibility forward fit approach is that
knowing the errors on visibilities $V(u,v; \epsilon )$ one
can readily propagate the errors to forward fit parameters of the
model in Equation (\ref{eq:Imodel}). Reliable error estimates
for images reconstructed with other algorithms
are currently unavailable.

\subsection{The shape of footpoints for a limb event}

Using HXR data from RHESSI we analysed a limb event on January 6th, 2004
(GOES M5.8 class). As shown previously by \citep{Kontar_etal08}, this event is ideally
suited for our analysis having two well separated footpoints: one bright
and a second much weaker footpoint. In addition, the location of the flare at the
limb greatly reduces albedo flux \citep{BaiRamaty1978,2010A&A...513L...2K},
so that the albedo correction \citep{Kontar_etal2006} becomes negligible.

\begin{figure}\center
\includegraphics[width=0.75\columnwidth]{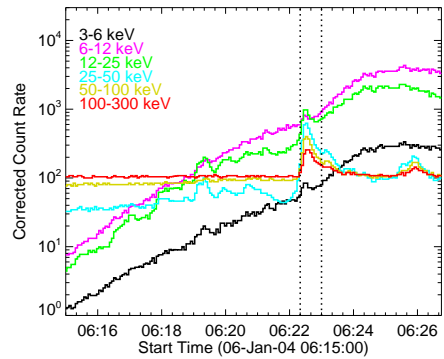}
\caption{
Lightcurves of the 6-Jan-2004 flare. The vertical dotted lines show the accumulation
time interval 06:22:20-06:23:00 UT which is used in the subsequent spectral and
imaging analysis. } \label{fig:lc}
\end{figure}
\begin{figure*}\center
\includegraphics[width=1.\columnwidth]{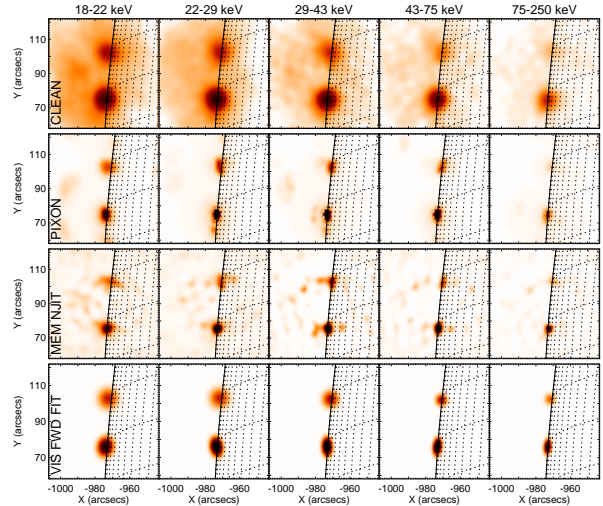}
\caption{HXR images of the January 6th 2004 flare (accumulation time interval 06:22:20-06:23:00
UT) in 5 energy ranges: 18-25, 25-29, 29-43, 43-75, 75-250 keV (energy grows from
left to right) using different algorithms: (from top row to bottom) CLEAN, PIXON,
MEM-NJIT, VIS forward fit. Image sizes are $64 \times 64$ with a pixel size of
$1\times 1$ arcseconds. Visibility forward fit images shown here were done using one elliptical
(strong southern footpoint) and circular (weak northern footpoint) Gaussian.}
\label{fig:images_jan06}
\end{figure*}

The flare occurred at the eastern limb near $(-975,75)$ arcseconds from the disk center
at $\sim 06:22$ UT (Figure \ref{fig:lc}). It was imaged during the time of peak emission $>50 $~keV
06:22:20-06:23:00 UT indicated by the vertical dotted lines in Figure \ref{fig:lc}, using
four different image algorithms (see Figure \ref{fig:images_jan06}): Clean
\citep{hurford2002}, MEM NJIT \citep{schmahl2007}, PIXON
\citep{PinaPuetter1993,Metcalfetal1996} and visibility forward fit
\citep{hurford2002,schmahl2007}. The resulting images in five energy bands covering
the nonthermal emission are shown in Figure \ref{fig:images_jan06}. Each image was
made using the front segments of detectors 2 to 7. Grid 1 with the highest spatial
resolution had no significant signal and grids 8-9 are too coarse for our
flaring region. Previously the flare was imaged using ten energy
bins \citep{Kontar_etal08} and simple circular gaussian fit
but this was reduced to five wider bins in this paper
to improve signal to noise. Figures \ref{fig:images_jan06},\ref{fig:ff_fit}
demonstrate that the brighter source has an elliptical shape
at various energies, so an elliptical Gaussian
could be used as natural X-ray distribution model
(Figure \ref{fig:ff_fit}).

Comparing the different algorithm results we find that CLEANed
images have systematically larger sizes than the other algorithms.
This is related to the fact that CLEAN images are determined
by the user choices for analysis (clean beam size) and not the requirements
of the data and hence should be used with great care to measure source sizes.
\footnote{Reduction of the CLEAN beam size by $~1.7$ produces images with spatial
characteristics similar to other algorithms. The current version of clean does
not have a robust procedure to determine the CLEAN beam size. Note that this correction
is only applicable for this particular event and cannot be used universally.}
MEM NJIT has produced smaller source sizes, which could be the tendency
of the algorithm to over-resolve sources \citep{schmahl2007}.
PIXON \citep{PinaPuetter1993,metcalf1995} gave source sizes similar
to those of X-ray visibility forward fit. \citet{dennis2009} have
also analysed this event and confirmed the finding of \citet{Kontar_etal08}. We
choose to forward fit a circular Gaussian source for the northern footpoint and
an elliptical Gaussian source (Equation \ref{eq:Imodel}) for the southern footpoint
to the visibilities (Equation \ref{eq:vis}), the image
shown is a reconstruction of the fit results. These fits are shown in Figure
\ref{fig:ff_fit} and will be discussed in detail in \S\ref{sec:fit}.
Assuming two elliptical sources, the weaker source forward fit
parameters have rather large error bars suggesting that Northern footpoint
is not sufficiently well-constrained by the data to be fitted
as an elliptical source. In addition, at the energies above
$40$~keV the weak source is indistinguishable from circular.

The comparison of the images and visibility fit results
in Figure \ref{fig:images_jan06} shows that visibility
forward fit gives images similar to the ones inferred in
other algorithms, although there are differences pointed out above.
Despite the differences between the algorithms, all image reconstruction
algorithms show that the southern footpoint has a) a clear elliptical shape, b) the
shape of the source becomes more elliptical with growing energy c) the size of the
source decreases with energy. The northern footpoint is also getting smaller with
energy similar to the southern footpoint \citep{Kontar_etal08,dennis2009}, but due to
lower count rate in the source we cannot reliably measure the shape of this source. In
addition, since the northern source is not seen above $\sim 100$~keV, it could be
partially occulted or have stronger magnetic convergence \citep{Schmahl_etal2006}
with the energetic electrons precipitating less to dense layers
of the chromosphere, producing a fainter footpoint.

\subsection{Characteristic sizes and foot-point locations}\label{sec:fit}

We focus on the brighter southern footpoint fitted with an elliptical gaussian
(Equation \ref{eq:Imodel}) as more
spatial information can be accurately recovered compared
to the northern footpoint.
This is a more realistic interpretation of the footpoint shape (see Figure
\ref{fig:images_jan06}) compared to the previously used circular
fit \citep{aschwanden2002,Kontar_etal08}.

\begin{figure}\center
\includegraphics[width=.45\columnwidth]{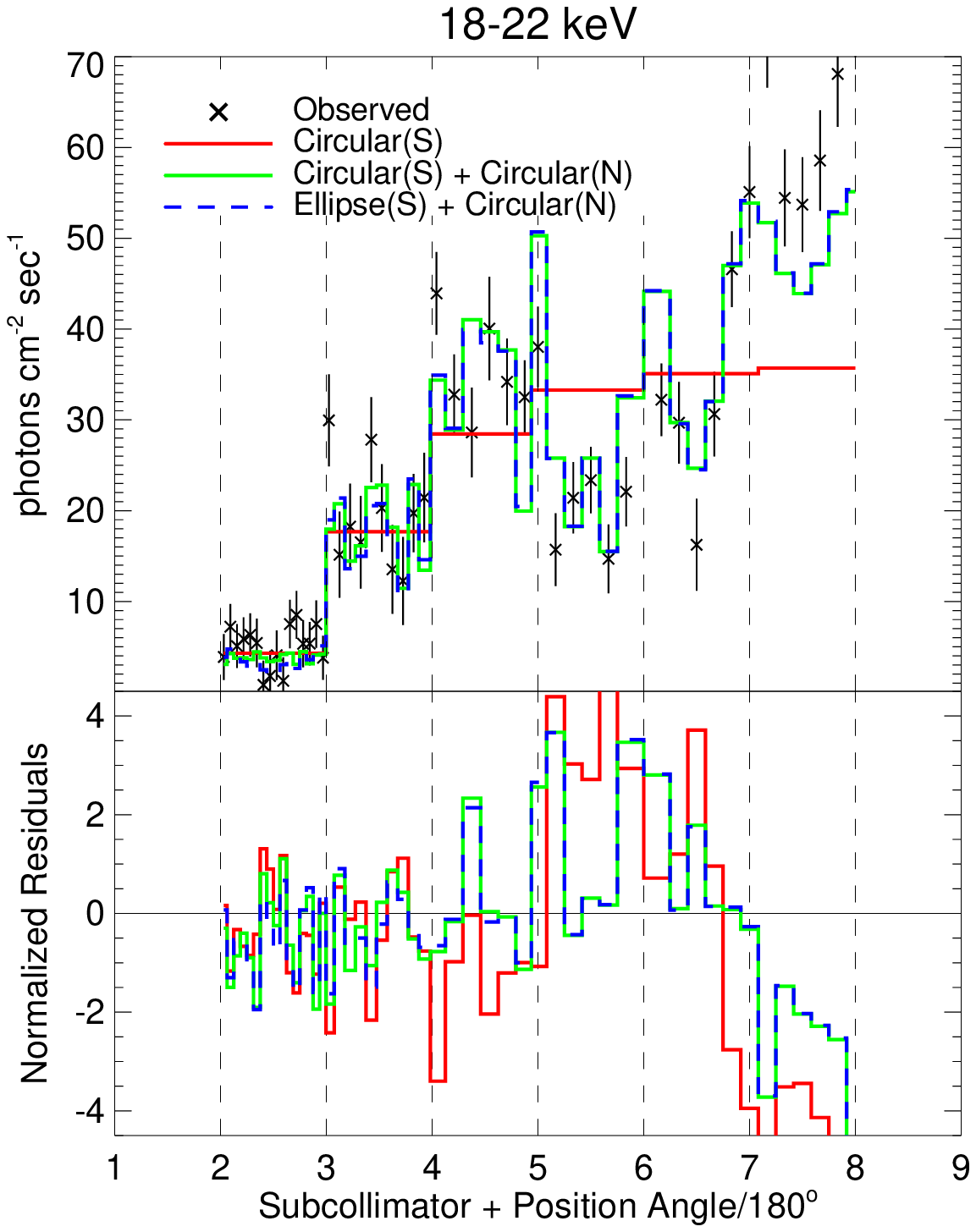}
\includegraphics[width=.45\columnwidth]{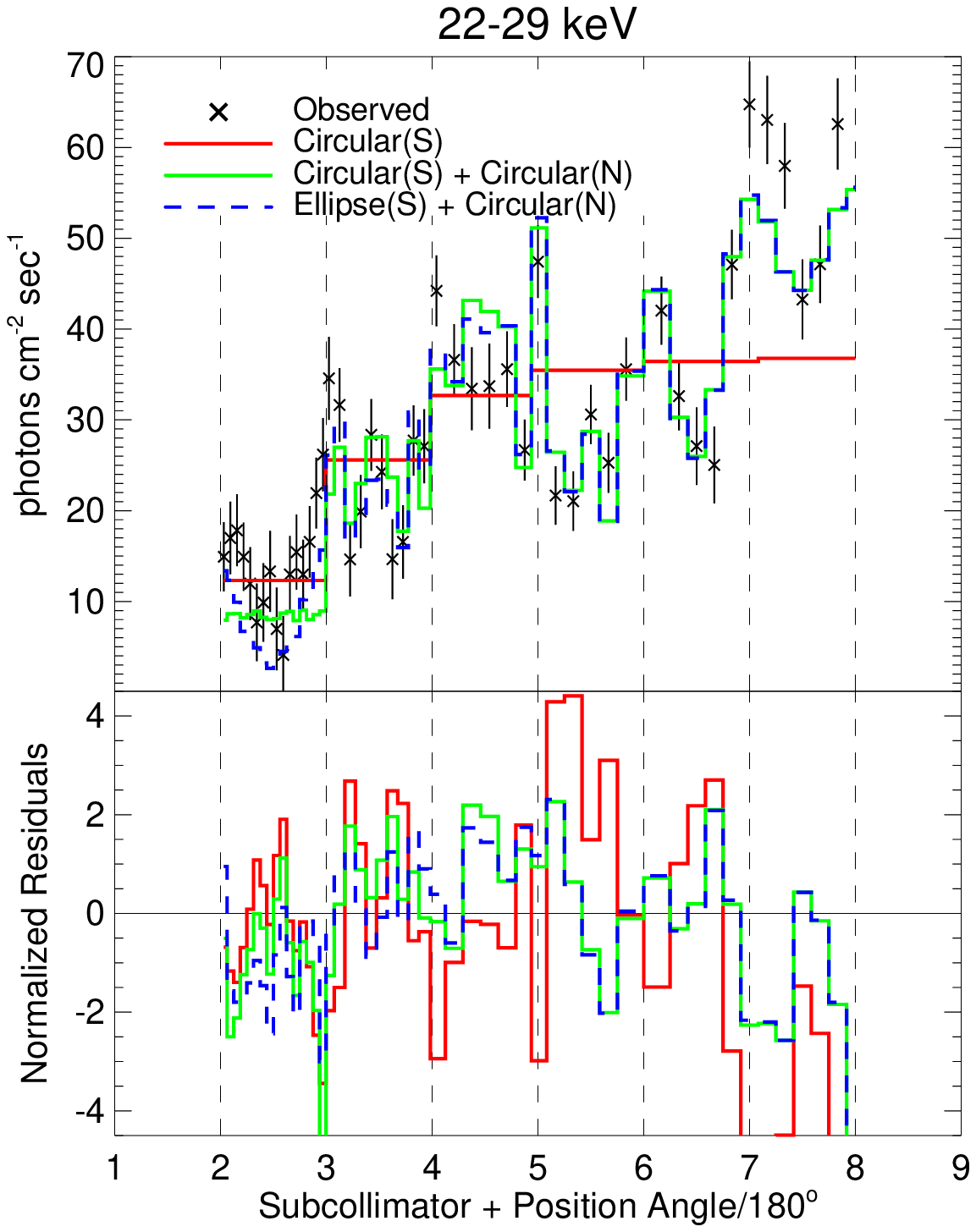}\\
\includegraphics[width=.45\columnwidth]{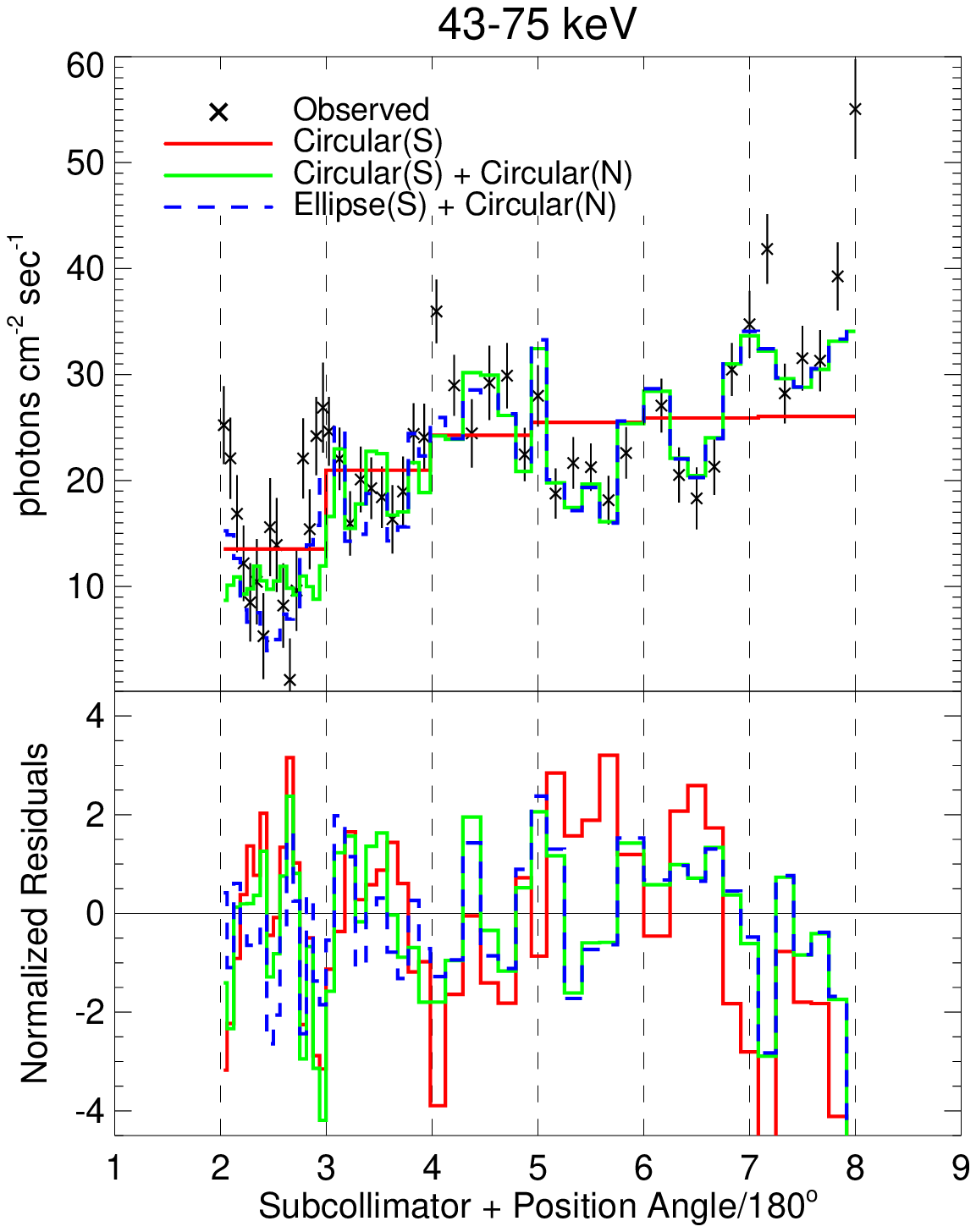}
\includegraphics[width=.45\columnwidth]{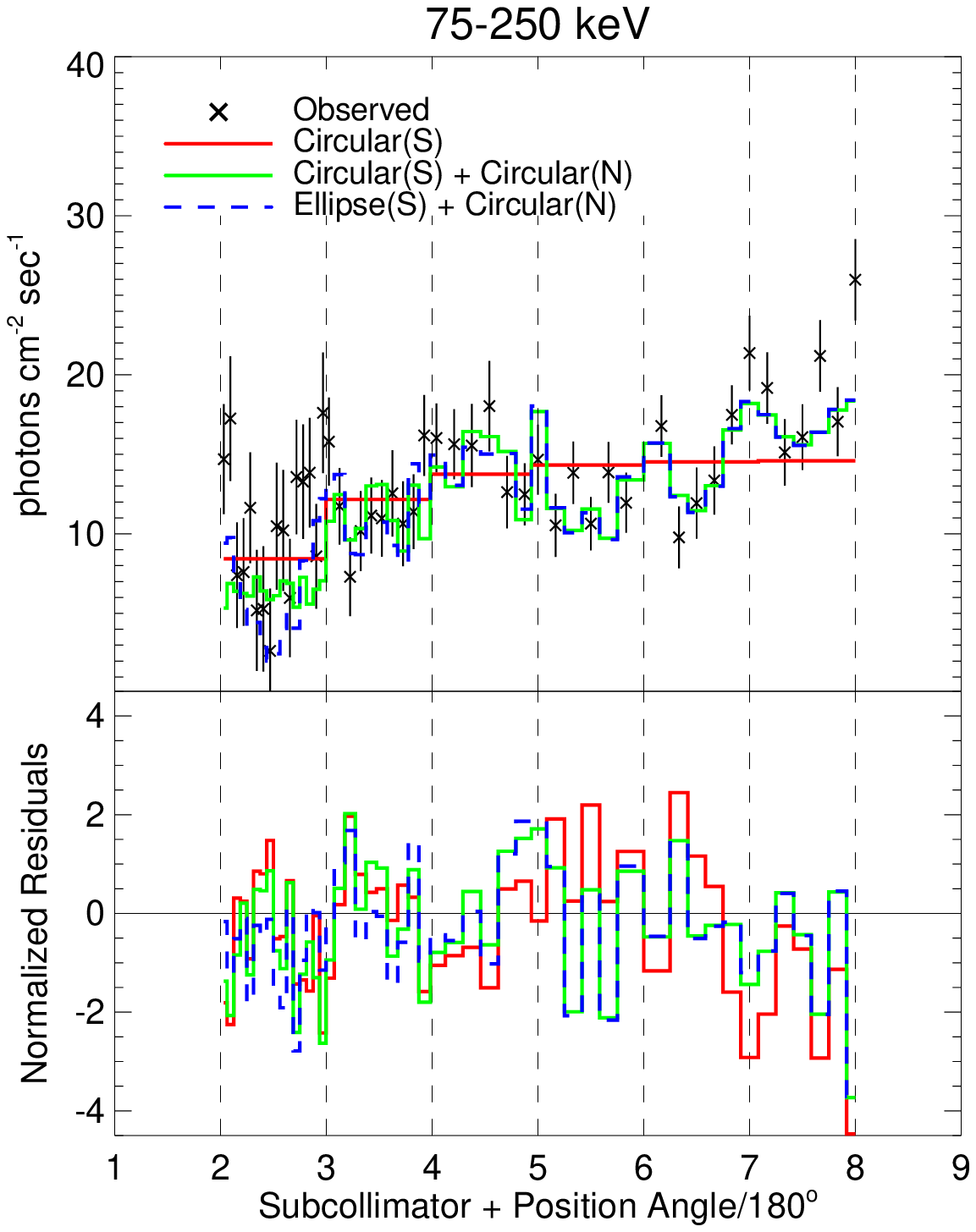}
\caption{Observed X-ray visibility amplitudes (crosses with error bars) as a function
of subcollimator (2-7) and position angle between $0$ and $180^o$ of the grids in the stated
energy range. The red line shows the fitted model
using single circular source, the green line shows two circular Gaussian fits, and the blue dashed line shows an elliptical (southern footpoint) and circular (northern footpoint) Gaussian.
The bottom panels show normalised residuals for the corresponding fits.}
\label{fig:ff_fit}
\end{figure}

Each forward fit to X-ray visibilities using an elliptical gaussian produces 6
parameters given by Equation (\ref{eq:Imodel}): positions $x_0(\epsilon)$ and $y_0(\epsilon)$
of the X-ray flux maximum (often called centroid position), full width half maximum $FWHM(\epsilon)$,
eccentricity $e(\epsilon)$ and position angle $\theta(\epsilon)$, along with error values
for each of the parameters. $x_0(\epsilon)$ and $y_0(\epsilon)$ are measured from disk centre and
the position angle $\theta(\epsilon)$ is the angle between the North-South line and
the semi-major axis of the ellipse. Multiple sources are fitted simultaneously and we fit
the weaker northern source with a circular Gaussian. Both sources can be fitted with
elliptical sources producing similar results for the southern footpoint but highly
inaccurate results for the northern footpoint. The visibility amplitudes and fits as a function of
RMC and spacecraft roll angle are shown in Figure \ref{fig:ff_fit}. We used
between 6 (course grids) and 12 (fine grids) visibilities (spatial Fourier components)
(Figure \ref{fig:ff_fit}). The single circular Gaussian fit shows the largest amplitudes
of normalised residuals. Two circular Gaussian fit has smaller amplitudes,
but larger than the fit using an elliptical and circular Gaussian fits.
The circular plus elliptical fits adequately reproduce the measured photon
flux for various roll bins and collimators. We note that
at the lowest energies $18-22$ keV, both two circular and elliptical plus
circular Gaussians give almost identical results. Indeed, both footpoints at $18-22$ keV
look symmetrical (Figure \ref{fig:images_jan06}). The largest deviations of the fits
from the data is found in the coarsest grid and at the lowest energy (Figure \ref{fig:ff_fit}),
which could be caused by the large scale source ($\gtrsim 36''$),
probably softer X-ray emission from the loop.

Using the $(x_0(\epsilon),y_0(\epsilon))$ ``centroid positions'' of the source for each
energy range, the radial height of the source from the solar centre can be readily
determined for every energy using:
\begin{equation}
R(\epsilon)=\sqrt[]{x_0(\epsilon)^2+y_0(\epsilon)^2}.
\end{equation}
\label{eq:R_eps} The semi-major and semi-minor axes of our elliptical gaussian fit,
$a(\epsilon)$, $b(\epsilon)$ respectively are related to the FWHM:
\begin{equation}
\mbox{FWHM}(\epsilon) = \sqrt{a(\epsilon)b(\epsilon)},
\label{eq:FWHM}
\end{equation}
and to the eccentricity by:
\begin{equation}
e(\epsilon)=\sqrt{1-b(\epsilon)^2/a(\epsilon)^2}.
\label{eq:ecc}
\end{equation}
Equations (\ref{eq:FWHM}) and (\ref{eq:ecc}) can be solved to find
\begin{equation}
 a(\epsilon)=FWHM(\epsilon)(1-e(\epsilon))^{-1/4}, \quad
b(\epsilon)=
FWHM(\epsilon)(1-e(\epsilon))^{1/4}
\end{equation}

\begin{figure}\center
\includegraphics[width=0.65\columnwidth]{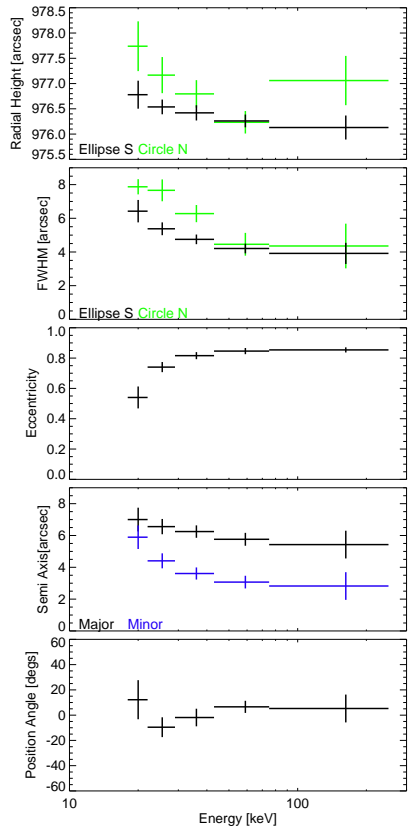}
\caption{
X-ray visibilities forward fit parameters as a function of energy: radial distance of the sources, FWHM of the sources, eccentricity,
semi-minor and semi-major axes for the northern footpoint, and position angle
(the angle between semi-major axis and North-South direction).}
\label{fig:vis_fits_jan06}
\end{figure}
As the analysed flare is right on the Eastern limb and close to the solar equator,
$a(\epsilon)$ and $b(\epsilon)$ correspond to the source sizes
parallel (width) and perpendicular (vertical extent) to the solar surface. The results of the X-ray visibility
forward fit parameters are summarised in Figure
\ref{fig:vis_fits_jan06}. They again show the trend seen in the images (Figure
\ref{fig:images_jan06}) of decreasing height and source size with energy. We can
also see that the source becomes more elliptical at higher energies, starting with
$e\sim0.5$ for 18-22 keV but increasing to $e\sim 0.85$ for 75-250 keV. The circular
Gaussian source fitted to the northern footpoint also shows the general trend of
decreasing source height and FWHM at higher energies but with considerably
larger errors due to this source being weaker.

\subsection{Height of X-ray sources above the photosphere}

Let us consider the evolution of the electron flux spectrum in the chromosphere
$F(E,s)$ along magnetic field lines $s$ using purely collisional transport and ignoring
collective effects and effects connected with the magnetic mirroring
\citep{brown2002}. In this approximation the electron flux spectrum can be written
\citep{brown1971}
\begin{equation}\label{eq:sol_F}
    F(E,s) = F_0(E _0)\frac{E}{E_0}
\end{equation}
where $F_0(E)$ is the injected spectrum of energetic electrons, taken to be a
 powerlaw of $F_0(E)\propto E^{-\delta}$ and
\begin{equation}\label{eq:sol_E0}
    E_0(E,s)^2=E^2 + 2K\int_{0}^{s}n(s')ds'
\end{equation}
where $K=2\pi e^2 \ln\Lambda$, $\ln\Lambda$ is the Coulomb logarithm, $e$ is the
electron charge. The chromosphere below the transition region can be conveniently assumed to be neutral
\citep{brown1973,kontar2002,2009ApJ...705.1584S} therefore $\ln\Lambda = \ln\Lambda _{eH} = 7$
\citep[e.g.][]{brown1973,emslie1978}.

The X-ray flux spectrum emitted by the energetic electrons in a magnetic flux tube of
cross-sectional area $A$ and observed at 1AU is given as
\begin{equation}\label{eq:sol_I}
    I(\epsilon,s)=\frac{1}{4\pi R^2}An(s)\int_{\epsilon}^{\infty}F(E,s) \sigma
(E,\epsilon)dE,
\end{equation}
where $\sigma (E,\epsilon)$ is the isotropic bremsstrahlung cross-section, $R$ is the
Sun-Earth distance, $A$ is the cross-sectional area of the loop. The X-ray flux spectrum
expressed by Equation (\ref{eq:sol_I}) has a maximum or equivalently $d
I(\epsilon,s)/ds=0$ for every energy $\epsilon$ because of the growing density along
electron path and simultaneously decreasing electron flux due to collisions
\citep{brown2002,aschwanden2002}.

\begin{figure}\center
\includegraphics[width=0.8\columnwidth]{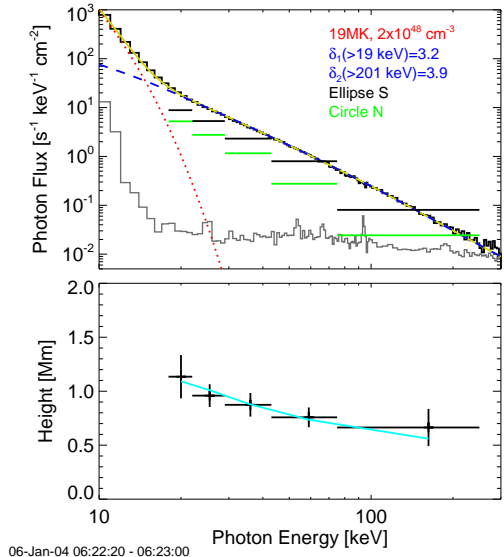}
\caption{
Upper panel: X-ray spectrum of January 6, 2004 flare (solid black histogram) for the
accumulation time 06:22:20-06:23:00 UT (see Figure \ref{fig:lc}) and forward fitted
spectrum (yellow line). The photon spectrum was fitted using a thermal (red dotted
line) plus thick-target model (blue dashed line). The gray histogram shows the
background level. Horizontal black and green lines show the photon flux spectrum from the southern and northern footpoints as found in X-ray visibility forward fits (Figure \ref{fig:images_jan06}) respectively. The lower panel indicate the results of the
forward fit of the source height with collisional model. The fit results are
$R_0=975.2''$, and $h_0=155$ km. } \label{fig:spectr}
\end{figure}

Assuming a hydrostatic density profile of
\begin{equation}
n(h=r-r_0)=n_0\exp{\left(\frac{-(r-r_0)}{h_0}\right)},
\end{equation}
where $r$ is the radial distance from the Sun centre, the photospheric density
$n_0=1.16\times 10^{17}$~cm$^{-3}$ [fixed value \citep{vernazza1981}], and $r_0$
is the reference height, we can find these two free parameters
$h_0$ and $r=r_0$ by forward fitting the measured radial distance of maxima (Figure
\ref{fig:spectr}, bottom panel) to the model predicted maxima by the derivative of
equation \ref{eq:sol_I}. The height of the sources can be found by subtracting the
reference height, $\mathit{r_0}$, from the radial measurements:
\begin{equation}
h(\epsilon)=r(\epsilon)-r_0.
\end{equation}
To calculate the reference height we
assumed the density at the photospheric level to be known \citep{vernazza1981}.
This helps to remove substantial uncertainties related
to the reference height of the previous studies
\cite[c.f.][]{aschwanden2002,Liu_etal2006,Mrozek06}.

We find a spectral index of $\delta=3.2$ from the spatially integrated spectrum,
shown in the top panel of Figure \ref{fig:spectr}. Forward fitting using a
hydrostatic density profile gives density scale height of $h_0=155\pm 30$~km and
reference height $r_0=975.2\pm0.2''$. From only fitting a circular gaussian to the
southern footpoint, instead of elliptical to southern with circular to the northern as
done in this paper, it was previously  found that $h=140 \pm 30$~km and
$r_0=975.3\pm0.2''$ \citep{Kontar_etal08}.

\subsection{Vertical extent of the footpoint}

\begin{figure}\center
\includegraphics[width=0.8\columnwidth]{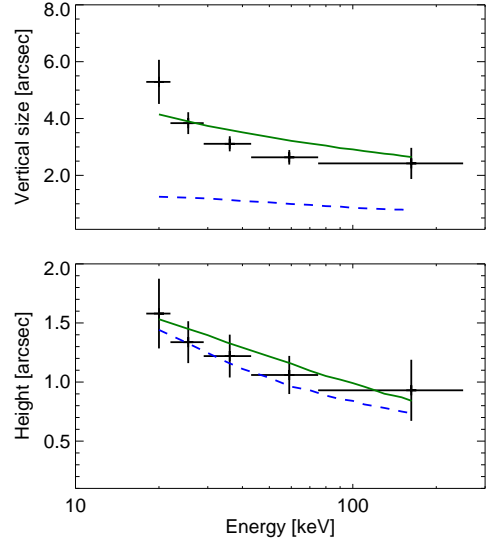}
\caption{Top panel: FWHM vertical size of the footpoint source (black crosses)
as a function of energy; FWHM size given by the
thick-target model (Equation \ref{eq:sol_I}) (blue dashed lines),
multi-thread chromosphere (green lines).
Bottom panel: height of the maximum X-ray emission (black crosses)
versus energy.}
\label{fig:dens_pert}
\end{figure}

The characteristic size of the source in the vertical direction (semi-minor axis in our fit)
can be straightforwardly estimated using the collisional think-target model
\citep{brown1971} to find the FWHM size from Equation \ref{eq:sol_I}
using the density profile found in the previous section.
Comparing the measured vertical FWHM size and the prediction of the length
from the thick-target model,
we see that the measured extent is around $3-6''$ and is
$3-6$ times larger than the theoretical width (Figure \ref{fig:dens_pert}).
The discrepancy is substantial and cannot be explained in terms of the assumed
X-rays source model or error bars.

\begin{figure}\center
\includegraphics[width=0.8\columnwidth]{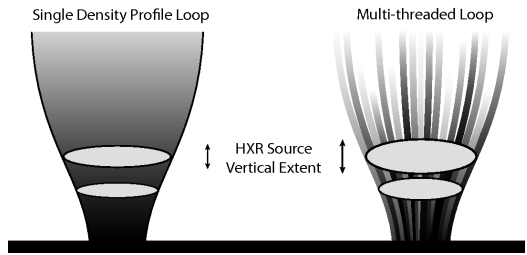}
\caption{Cross-sectionally uniform (left) and multi-threaded (right) chromospheres.
The multi-thread chromosphere leads to the larger vertical sizes
of X-ray sources as observed by RHESSI.}\label{fig:multi_thread}
\end{figure}

There are a few plausible explanations which can be given
for the observed vertical FWHM size. One is that the structure
of the chromosphere may not be uniform over the footpoint cross-section
and the footpoint is the ensemble of many thin
threads (Figure \ref{fig:multi_thread}) as often seen
in high resolution optical images \citep{LinEngvold_etal2005,DePontieu_etal2007,Berkebile-Stoiser_etal2009}.
Moreover, it is likely that the deposition of electron energy can lead to heating
of the chromosphere upwards with the resulting expansion changing
the density structure \citep[e.g.][]{Liu_etal2009}.

Let us consider a simple chromospheric model in which the hydrostatic density scale
height, $h_T$, is determined by the temperature of the chromosphere
$h_T=k_bT/(\mu m_pg)$ where $T$ is the temperature,
$k_b$ is the Boltzman constant,
$m_p$ is the proton mass, $\mu \simeq 1.27$ is the mean molecular
weight \citep[e.g.][]{aschwanden2002}, and $g \simeq 2.74\times 10^4 $~cm~s$^{-2}$
is the solar gravitational acceleration. Thus, the measured density scale height
of $\sim 155$~km implies an average chromospheric temperature of $\sim 6500$~K.
However, the solar atmosphere is not a uniform media but instead is
manifested in thin threads $(<0.3)''$ of filaments \citep{LinEngvold_etal2005},
sub-arcsecond dynamic fibrils \citep{DePontieu_etal2007} and fine structure
of microflares \citep{Berkebile-Stoiser_etal2009} in the chromosphere
with different density profiles in each thread (Figure \ref{fig:multi_thread}).
Following the multi-thread model for the magnetic loop in the chromosphere,
we assume that the energetic electrons propagate along different
threads with varying density profiles and temperatures
in the range from $2200$~K up to $20000$~K, corresponding
to density scale heights between $\sim 50$~km and $\sim 500$~km.
The average hard X-ray flux from many thin threads
will be the measured X-ray distribution. Averaging X-ray emission
from a hundred thin threads with temperatures drawn randomly
from the above range, we successfully reproduce
the observed vertical
sizes (Figure \ref{fig:dens_pert}).

\section{Discussion and Conclusions}

Using X-ray visibility forward fits, we inferred not only the characteristic sizes and
positions but the shapes of HXR sources. The January 6th 2004 event indicates an
overall decrease in size of the source and increase of the source ellipticity with
energy. The FWHM of the southern source decreases from $\sim6.5''$ down to around
$\sim 4''$ while the ellipticity of the source grows from $0.5$ up to $0.9$. The source
is elongated along the limb as evident in nearly zero angle between semi major-axis
and the limb, such orientation of the source is observed for all energy ranges. The
vertical extent of the source is decreasing by a larger fraction (from $6''$ down to
$3''$) than the horizontal size (from $7''$ down to $5.5''$) leading to larger
elongation of the source along the limb. Hence the FWHM of the magnetic flux tube
containing energetic electrons (semi-major axis) changes
from $\sim 9.7$~Mm at height $\sim 1.1$~Mm down to $\sim 7.6$~Mm
at $0.6$~Mm above the photosphere. The northern footpoint is fainter, but
shows a similar trend: the higher energies appear at low heights and the size of the
source is decreasing with energy suggesting convergence of the magnetic field
lines along which electrons propagate. Using X-ray visibilities we also re-analyzed a flare
that occurred on February, 20th, 2002 that was previously studied by forward fitting
time-modulated lightcurves \citep{aschwanden2002}. Although this flare
is rather weak, we found similar results: the higher energy sources appear
at lower heights. The uncertainties are larger than in the January 6th, 2004 event
but the density model proposed by \citet{aschwanden2002} is within our error
bars. We also note that the size of the sources in the February 20th flare decreases with
energy similar to the event on January 6th.

Analysing HXR emission from footpoints we found that various imaging algorithms
(PIXON, Visibility Forward Fit, CLEAN, MEM-NJIT) give generally similar
results for spatial distributions of X-ray footpoints. Although RHESSI imaging algorithms
can be adjusted by the parameter choice and hence X-ray images could be somewhat
altered, there is a general trend for the algorithms. CLEAN with default set of parameters
has a tendency to provide larger sources while MEM-NJIT tends to over-resolve X-ray sources.
PIXON and visibility forward fit show very similar results. Visibility forward
fit allows us to study sub-arcsecond distribution
of hard X-ray sources in suitably orientated bright flares. Due to systematic
differences in sizes we obtained for January 2004 and February, 2002 flares, we suggest
that CLEAN and MEM-NJIT should be used for source size/shape measurements
of X-ray sources with extreme caution.

The northern footpoint could be partially occulted
as the highest energy photons come predominantly from the southern footpoint, but
we note that this bright footpoint is unlikely to be occulted.
As pointed out by G. Hurford\footnote{Presentation at 9th RHESSI workshop in Genoa
http://sprg.ssl.berkeley.edu/~krucker/genoa/position/XrayLimb-Genoa.ppt},
a source partially subtended by the solar
disk should have a sharp edge, where the brightness of the source will drop
from maximum to zero over rather small radial distance. The derivative
of the source brightness in the $x$-direction (perpendicular to the limb),
$\partial I(x,y)/\partial x$ will have a maximum at the limb,
where $I(x,y)$ has a sharp drop. It is evident from Equation (\ref{eq:vis})
that the corresponding visibilities, $2\pi iuV(u,v)$,
should have a well pronounced maximum, which should be evident
in the measured amplitudes of visibilities. Specifically, the finer grid RMCs
should show large amplitudes when the grids are parallel to limb, i.e. the visibility
amplitudes are much larger at the phase angles $~0^o$ and $~180^o$, which is not evident
in the event under study (cf Figure \ref{fig:ff_fit}).
To make the discussion more complete, we note that the line of sight
effects cannot be definitively ruled out. The footpoints might be projections
of two rather long flare ribbons viewed almost parallel to the line of sight,
so that the vertical extension is the projection of different height.
Finally we should note that if the occultation height is small, $\lesssim0.5''$,
our height measurements are lower limits, but the major conclusion about the
vertical extend is the same. If, though unlikely, the lower part of the loop
(footpoints) is occulted, these observations provide an interesting question
to the flare models as to why the sources sizes decrease with energy
and the higher energy sources appear lower and not at the lowest visible
location where the density is the highest.

Our measurements show that while the locations of the maxima of X-ray emission
are consistent with simple collisional transport in single density scale height
chromosphere, the vertical sizes do not agree
with the assumption of field aligned electron transport.
The vertical extent of X-ray sources is 3-6 times larger
than in the purely collisional model in single-density-scale-height chromosphere.
However, a chromospheric model involving multiple density threads within the flux tube
of a footpoint can explain both the position of the maximum and the vertical
size of the sources.
We note that pitch angle scattering due to Coulomb
collisions is likely to be insufficient to produce so strong expansion.
The X-ray source size increase due to collisional pitch angle scattering
will be about a quarter of the electron stoping depth
for initially field aligned electrons \citep{Conway2000}.
However, strong non-collisional scattering or wave-particle
interactions \citep[e.g.][]{Hannah_etal2009} might boost
the vertical source sizes to the measured value
and hence cannot be excluded and will also be consistent with
lack of downward anisotropy found in X-ray flare emission
\citep{KontarBrown06mirror,Kasparova_etal2007}.
We note that the adopted model does not account for
the magnetic field and its effects on particle transport,
which could lead to larger source sizes and is subject
of additional modeling.



\acknowledgments

The authors are thankful to Gordon Hurford for insightful suggestions and Richard Schwartz
for referee comments. This work was supported by a STFC rolling grant (IGH, EPK),
STFC/PPARC Advanced Fellowship (EPK), and the Leverhulme Trust (MB,EPK). Financial
support by the European Commission through the SOLAIRE Network (MTRN-CT-2006-035484) is
gratefully acknowledged. The work has benefited from ISSI, Bern visitor programme.

\bibliographystyle{apj}
\bibliography{ms}





\end{document}